\documentclass[final,3p,times,twocolumn]{elsarticle}    

\usepackage{hyperref}
\usepackage{amssymb}
\usepackage{lipsum}

\newcommand{\um}{\,µm} 
\newcommand{\us}{\,µs} 

\newcommand{\allpix}{Allpix$^2$} 

\journal{Nucl. Instrum. Methods Phys. Res. A}

\begin{document}

\begin{frontmatter}



\title{Simulations and Performance Studies of a MAPS in 65\,nm CMOS Imaging Technology}



 \affiliation[desy]{organization={Deutsches Elektronen-Synchrotron},
            addressline={Notkestraße 85},
            city={22607 Hamburg},
            postcode={},
            country={Germany}}

\affiliation[bonn]{organization={University of Bonn},
            addressline={Regina-Pacis-Weg 3},
            city={53113 Bonn},
            postcode={},
            country={Germany}}

\affiliation[cern]{organization={Conseil Européen pour la Recherche Nucléaire},
            addressline={Esplanade des Particules 1},
            city={1211 Geneva 23},
            postcode={},
            country={Switzerland}}

\affiliation[hamburg]{organization={University of Hamburg},
            addressline={Mittelweg 177},
            city={20148 Hamburg},
            postcode={},
            country={Germany}}

\affiliation[giessen]{organization={University of Giessen},
            addressline={Ludwigstraße 23},
            city={35390 Gießen},
            postcode={},
            country={Germany}}

\affiliation[wuppertal]{organization={University of Wuppertal},
            addressline={Gaußstraße 20},
            city={42119 Wuppertal},
            postcode={},
            country={Germany}}

\affiliation[mainz]{organization={Johannes Gutenberg University of Mainz},
            addressline={Saarstraße 21},
            city={55122 Mainz},
            postcode={},
            country={Germany}}

\author[desy,bonn]{Adriana Simancas\corref{mycorrespondingauthor}}
\cortext[mycorrespondingauthor]{Corresponding author}
\ead{adriana.simancas@desy.de}
\author[cern,hamburg]{Justus Braach}
\author[cern]{Eric Buschmann}
\author[desy]{Ankur Chauhan}
\author[cern]{Dominik Dannheim}
\author[desy,bonn]{Manuel Del Rio Viera}
\author[cern,giessen]{Katharina Dort}
\author[desy]{Doris Eckstein}
\author[desy]{Finn Feindt}
\author[desy]{Ingrid-Maria Gregor}
\author[desy]{Karsten Hansen}
\author[desy]{Lennart Huth}
\author[desy,bonn]{Larissa Mendes}
\author[desy]{Budi Mulyanto}
\author[desy,wuppertal]{Daniil Rastorguev}
\author[desy]{Christian Reckleben}
\author[desy,bonn]{Sara Ruiz Daza}
\author[desy,mainz]{Judith Schlaadt}
\author[desy]{Paul Schütze}
\author[cern]{Walter Snoeys}
\author[desy]{Simon Spannagel}
\author[desy]{Marcel Stanitzki}
\author[desy]{Anastasiia Velyka}
\author[desy,bonn]{Gianpiero Vignola}
\author[desy]{Håkan Wennlöf}

\begin{abstract}
Monolithic active pixel sensors (MAPS) produced in a 65\,nm CMOS imaging technology are being investigated for applications in particle physics. The MAPS design has a small collection electrode characterized by an input capacitance of $\sim$fF, granting a high signal-to-noise ratio and low power consumption. Additionally, the 65\,nm CMOS imaging technology brings a reduction in material budget and improved logic density of the readout circuitry, compared to previously studied technologies. Given these features, this technology was chosen by the TANGERINE project to develop the next generation of silicon pixel sensors. The sensor design targets temporal and spatial resolutions compatible with the requirements for a vertex detector at future lepton colliders. Simulations and test-beam characterization of technology demonstrators have been carried out in close collaboration with the CERN EP R\&D program and the ALICE ITS3 upgrade. TCAD device simulations using generic doping profiles and Monte Carlo simulations have been used to build an understanding of the technology and predict the performance parameters of the sensor. Technology demonstrators of a 65\,nm CMOS MAPS with a small collection electrode have been characterized in laboratory and test-beam facilities by studying performance parameters such as cluster size, charge collection, and efficiency. This work compares simulation results to test-beam data. The experimental results establish this technology as a promising candidate for a vertex detector at future lepton colliders and give valuable information for improving the simulation approach.
\end{abstract}



\end{frontmatter}




\section{Introduction}

Lepton colliders have been established as the highest-priority next collider by the European Strategy Update for Particle Physics~\cite{european_strategy}. Vertex detectors are an essential part of experiments at such colliders. They require simultaneous advances in material budget, granularity, and spatial and temporal resolution. Monolithic CMOS sensors are promising candidates given these requirements and have the advantage of cost-efficient mass production capabilities in commercial foundries. The TPSCo 65\,nm ISC imaging CMOS technology is currently being studied for applications in particle physics. Introducing this node size in high energy detectors will improve the in-pixel logic density and/or will allow for a reduction in the pixel pitch.

\textit{Monolithic active pixel sensors} (MAPS) can be produced with a small or a large collection electrode. This work explores the small collection electrode type MAPS motivated by the small intrinsic capacitance (in the order of fF) and, hence, a large signal-to-noise ratio. The activities are carried out within the context of the TANGERINE Project~\cite{Wennlof2022,Feindt2022,Simancas2023}, which aims to develop the next generation of silicon detectors for vertex-finding at future lepton colliders. To achieve this, the performance targets shown in Table~\ref{tab:requirements}, must be fulfilled. The developments are pursued in collaboration with the CERN EP R\&D program~\cite{EPreport} on technologies for future experiments and with the ALICE ITS3 upgrade~\cite{alice}. 

\begin{table}
\caption{Requirements for vertex detectors at lepton colliders. Derived from sources such as~\cite{clic_CDR}.}
\begin{tabular}{l c c c} 
 \hline
\textbf{Parameters} & \textbf{Requirements}\\
\hline
Material Budget         & \textless 1\% $X_0$ \\
Single-point Resolution & $\leq 3~\mu m$ \\
Time Resolution         & $\sim$ ns \\
Granularity             & $\leq 25~\mu m \times 25~\mu m$  \\
Radiation Tolerance     & $> 10^{11} n_{eq}/cm^2$ \\
 \hline
\end{tabular}
\label{tab:requirements}
\end{table}

The collaboration included a common foundry submission for test chips in the TPSCo 65\,nm CMOS technology, as well as device characterization and generic simulations validating the sensor design for appropriate performance. From this first submission, two test structures have been tested. The DESY Chip V1 (designed at DESY) features a Krummenacher charge-sensitive amplifier and was characterized in~\cite{Feindt2022}. The Analog Pixel Test Structure (APTS)~\cite{Deng2023,alice_apts} (designed at CERN) is a technology demonstrator with analog readout designed to characterize different sensor layouts and is part of the studies for the ALICE ITS3 upgrade.

This work presents the test-beam characterization of an APTS, a description of the simulation approach, and a first comparison between experiment and simulations.

\section{Sensor Layouts}

The sensor design consists of a thin high-resistivity p-doped epitaxial layer grown on a low-resistivity p-doped substrate. The n-well and p-well are, respectively, the collection implant and the structure that hosts the in-pixel electronics and shields them from the electric field of the active sensor region. This base design is the \textit{standard layout}~\cite{Senyukov2013}, characterized by a bulb-shape depleted volume around the collection implant. 

\begin{figure}[htb]
\begin{center}
	\centering 
	\includegraphics[width=0.5\textwidth]{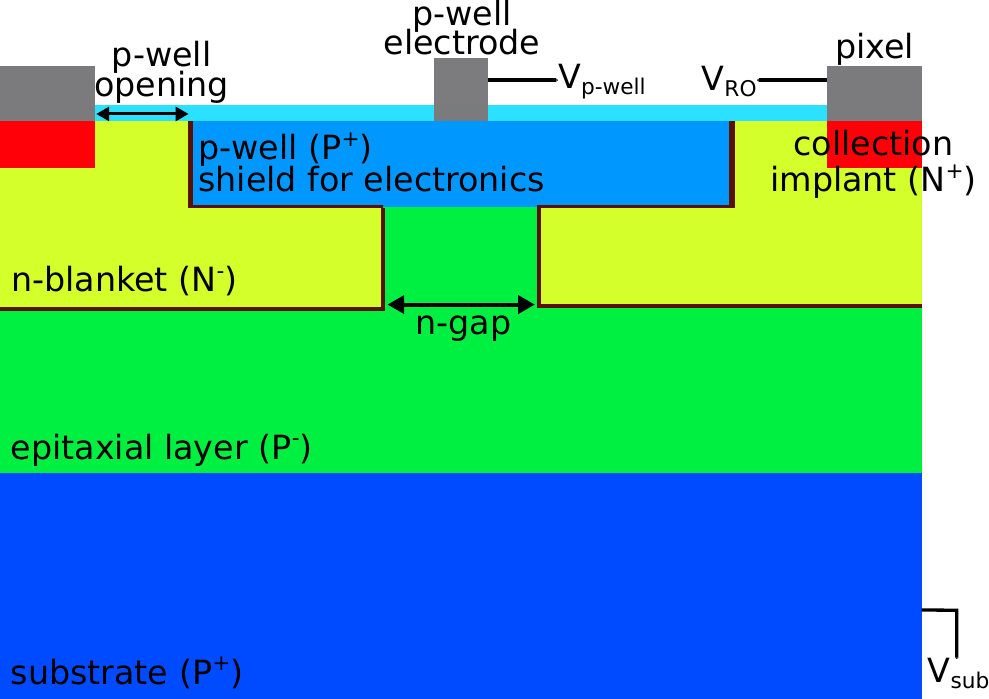}
	\caption{Scheme of the n-gap sensor design. Structures not to scale.}
 \label{fig:detailed_scheme_gap}%
 \end{center}
\end{figure}

Design modiﬁcations can improve the electric ﬁeld configuration inside the sensor. The \textit{n-blanket layout}~\cite{Snoeys2017} introduces a blanket layer of n-doped silicon in the p-type epitaxial layer, creating a deep planar pn-junction and enlarging the depleted volume of the sensor. However, this layout leaves an electric field minimum under the p-well at pixel edges and corners, leading to slow charge collection and possible efficiency loss in these regions~\cite{Muenker_thesis}. The \textit{n-gap layout}~\cite{Munker2019} corrects for this by introducing a gap in the n-blanket under the p-well. This produces a vertical pn-junction that generates a lateral electric field in the farthest position from the readout electrodes (pixel boundaries), pushing charges produced there toward the pixel center. This work is focused on the performance of the n-gap design, whose layout scheme is shown in Figure~\ref{fig:detailed_scheme_gap}. The electric field and depleted volume for each design can be found in~\cite{Simancas2023}.

These designs were originally developed in a 180\,nm CMOS imaging technology, and similar developments have been implemented in a 65\,nm CMOS imaging process as well~\cite{Snoeys_pixel22}.

The sensors are biased with a fixed positive voltage on the pixel electrodes and a negative bias voltage on the p-well electrode and the substrate. The voltage is the same on the p-well and substrate in these studies.

\section{Detector and Readout System}
\label{sec:prototype+DAQ}

The APTS~\cite{Deng2023} is a demonstrator designed for sensor characterization in the TPSCo 65\,nm CMOS technology. Each chip comprises a 6 × 6 matrix of square pixels, of which only the central 16 pixels are read out. The chips are available in all three sensor layouts, different doping variants, pixel geometries, and pixel pitches~\cite{alice_apts}. The device characterized in this work has an n-gap design with a 25\um{} pixel pitch; each pixel is DC coupled to the front-end electronics. Both in-pixel and periphery circuits contain two source-follower stages as buffered analog output.

The data acquisition setup comprises a custom chipboard for the APTS and the modular Caribou~system~\cite{caribou1,caribou2} consisting of open-source hardware, firmware, and software for prototype integration. Its key component is a System-on-Chip board (Xilinx Zynq) that runs the data acquisition software and firmware for powering, configuration, control, and readout of the prototype. A mezzanine board, the CaR board, provides current sources, voltage sources, and a physical interface to the chip. This board is used for efficiency studies and includes two 8-channel ADCs that sample the analog output signals of all pixels at 65\,MS/s.

Pulse injection measurements have been used to determine gain non-linearities and pixel-to-pixel variations. An absolute calibration of the gain curves was performed using the K-alpha line of an $^{55}$Fe source. The measured noise is about 30 electrons. More details on the DAQ system and calibration are reported in~\cite{Simancas2023}.

\section{Test-Beam Setup}

Test-beam studies allow for the characterization of new detector prototypes for particle physics applications under realistic conditions.

The test-beam setup consists of a MIMOSA26 telescope~\cite{mimosa}, composed of six planes and used for beam particle track reconstruction. The device under test (DUT), in this case, the APTS, is placed between the third and fourth telescope planes. The TelePix~\cite{telepix} detector is used as a trigger plane with a configurable acceptance window and is the last detector plane downstream. Finally, a Trigger Logic Unit (TLU)~\cite{tlu} manages the trigger signals of the setup to synchronize the data acquisition between all the devices.

The employed data acquisition software is EUDAQ2~\cite{eudaq2}, which controls the storage and synchronization of data from all systems. The data analysis framework for online monitoring and offline event building is Corryvreckan~\cite{corry}.

Two triggering schemes have been employed. To find the position of the DUT relative to the beam telescope, it was operated in self-triggered mode. To allow for unbiased efficiency measurements, it was triggered externally using TelePix, where a mask was applied to trigger on only a small region around the DUT.

The test-beam campaigns have been carried out at DESY-II~\cite{desy-tb}, with a 4\,GeV electron beam and a maximum beam particle rate of 5\,kHz.

\section{Test-Beam Characterization}
\label{sec:tb-characterization}

This section describes the reconstruction procedure of the test-beam data and discusses the results of cluster size and efficiency studies.

\subsection{Data Analysis}

When the DAQ receives a trigger, the respective waveforms are recorded. These are processed for each pixel during the data analysis; their amplitude is measured and transformed to charge using the calibration mentioned in Section~\ref{sec:prototype+DAQ}.

To calculate the amplitude, two regions are defined: the baseline region (1\us{} interval before the pulse starts) and the peak region (1.5\us{} interval around the expected pulse maximum). The baseline is obtained by averaging the values in the baseline region. The maximum of the pulse is taken from the peak region, and the amplitude is obtained by subtracting the baseline from the maximum value. A threshold is applied to define pixel hits. The studied thresholds are in the typical operating range, from 90 ($\sim 3\sigma_{noise}$) to 400\,electrons.

Individual hits on the same device that belong to the same particle interaction are grouped into a \textit{cluster} based on spatial vicinity. The employed clustering method reconstructs the cluster position and charge by defining a \textit{seed pixel} (pixel with the largest signal) and adding all the adjacent pixels with signal above thresholds. Then, the cluster position is calculated as the charge-weighted center of gravity.

The tracks of the beam particles are reconstructed using the telescope data; a fit is made using the General-Broken-Lines algorithm~\cite{gbl}, which considers the scattering of particles when passing through a material.

Finally, the clusters on the DUT are associated with reconstructed tracks within 30\um{} diameter window to study hit detection efficiency and cluster properties.

\subsection{Cluster Size}

The mean \textit{cluster size} is obtained by calculating the mean of the distribution in a cluster size histogram. It is highly dependent on its charge-sharing properties, which are regulated by the charge transport mechanism. In depleted areas, the movement of free charges is dominated by drift, usually directly towards the collection electrode. While in non-depleted areas, they move mainly by diffusion, producing a wider charge distribution and increasing charge sharing between the pixels.

\begin{figure}[htb]
	\centering
	\includegraphics[width=0.5\textwidth]{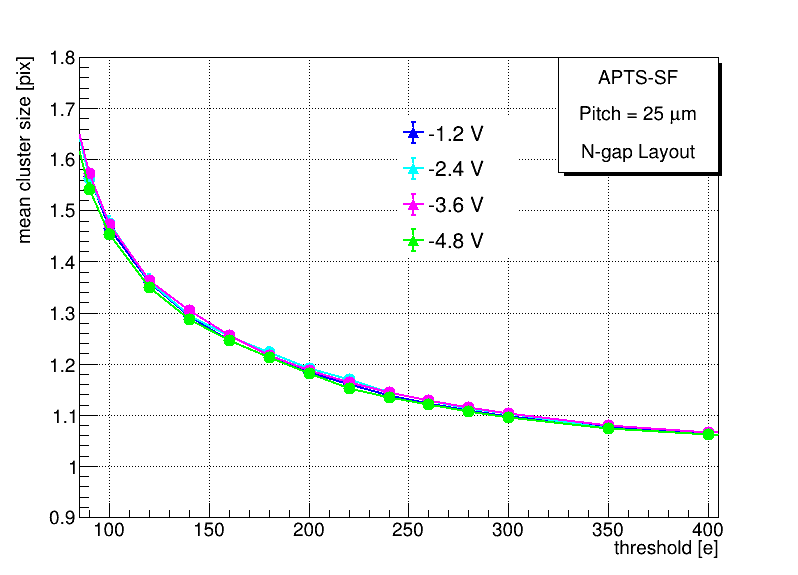}
	\caption{Mean cluster size as a function of the threshold for APTS in n-gap layout at different bias voltages, from -1.2 to -4.8\,V. Only statistical uncertainties are included.} 
	\label{fig:thScan_clust_bias_gap}%
\end{figure}

Figure~\ref{fig:thScan_clust_bias_gap} shows the mean cluster size as a function of the threshold, comparing the n-gap layout at different bias voltages. No significant difference is observed because the depleted volume in the sensor remains approximately constant for all bias voltages.

Within the operating thresholds, the n-gap design exhibits a mean cluster size ranging from 1.6 to less than 1.1~pixels. These results agree with the expectations from the design since the large depleted volume and the high lateral electric field in the edges constrain the charge-sharing effects. 

\subsection{Efficiency}

The \textit{detection efficiency} of the detector is calculated as the ratio of the associated clusters in the DUT and the reconstructed tracks within the acceptance window around the cluster center.

\begin{figure}[htb]
	\centering 
	\includegraphics[width=0.5\textwidth]{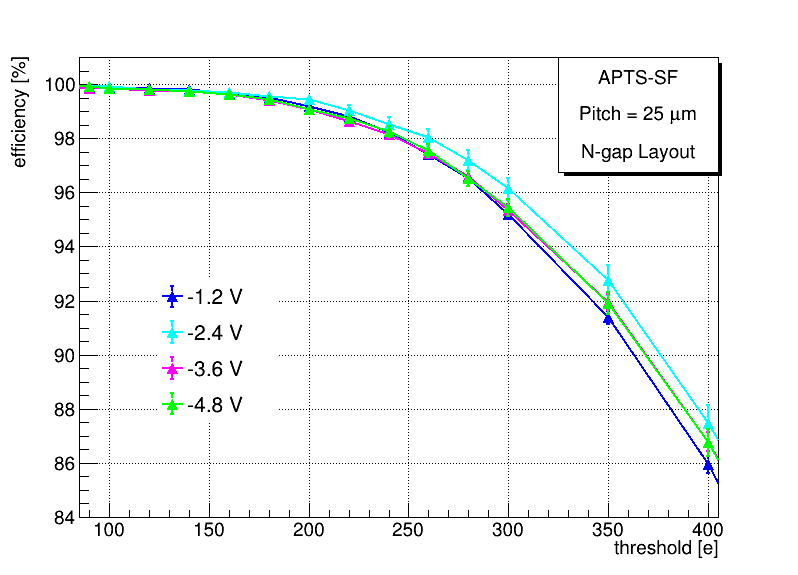}	
	\caption{Efficiency as a function of the threshold for APTS in n-gap layout at different bias voltages, from -1.2 to -4.8\,V. Only statistical uncertainties are included.} 
	\label{fig:thScan_eff_bias_gap}%
\end{figure}

Figure~\ref{fig:thScan_eff_bias_gap} compares the efficiency performance as a function of the threshold for the n-gap design at different bias voltages. The results are very similar since the depleted volume and the charge sharing in this layout are almost unchanged with the bias voltage. In the operating threshold range (above $3\sigma_{noise}$), the efficiency for the n-gap design starts at 99.9\% and falls to 86\%, and remains above 99\% until a threshold of 220\,electrons.

\section{Sensor Simulations}

The electric ﬁeld distribution in sensors greatly depends on doping proﬁles. In particular, MAPS with a small collection electrode have highly complex electric ﬁelds. Hence, sensor simulations are necessary to understand the inner workings of the detector. This work uses a combination of TCAD and Monte Carlo simulations to obtain precise electric fields and high statistic results for detector performance evaluations~\cite{tcad+mc}.

\subsection{TCAD Simulations}

The TCAD simulations carried out within the TANGERINE project are based on fundamental principles of silicon detectors and employ generic doping proﬁles as described in the following. A 3D structure is created with geometrical operations and analytic doping profiles following the scheme shown in Figure \ref{fig:detailed_scheme_gap}.

\textit{Quasi-stationary simulations} were performed to model the electric fields of the studied designs. These simulations aim to understand the effect of design changes and provide input for optimization of the design and the operational parameters. This was achieved by scanning over different geometrical and operational parameters of the sensor, such as p-well opening and bias voltage, and observing the behavior of the electric field, the lateral electric field strength, as well as the depleted volume. Finally, the parameters that reproduce the expected physical behavior (similar to previous studies on other technologies~\cite{Senyukov2013,Snoeys2017,Munker2019}) are selected to derive electric fields for subsequent Monte Carlo simulations. More details on this simulation approach can be found in~\cite{Simancas2023}.


\subsection{Monte Carlo Simulations}

Monte Carlo simulations are employed to model the full response of a detector. This is achieved with the modular framework \allpix~\cite{Spannagel2018}, developed for Monte Carlo simulations of semiconductor radiation detectors. The results of these simulations allow for direct comparison with experimental data.

The simulated detector structure consists of a matrix of square pixels with a 25\um{} pitch and a sensor thickness of 50\um{} (including epitaxial layer and substrate). The electric field for each pixel cell is imported from the TCAD simulation described above. A 4\,GeV electron beam is used as a particle source to compare with the test-beam results. The data processing is analog to the one described in Section \ref{sec:tb-characterization} for test-beam data. The most significant observables that can be obtained are detection efficiency, cluster size, spatial resolution, and charge collection. Initial results from Monte Carlo simulations using generic TCAD fields have been reported in~\cite{Wennlof2022}, and the most recent results on efficiency and charge collection are compared with test-beam data in the following section.

\section{Comparing Data and Simulation}

\begin{figure}[htb]
	\centering 
	\includegraphics[width=0.5\textwidth]{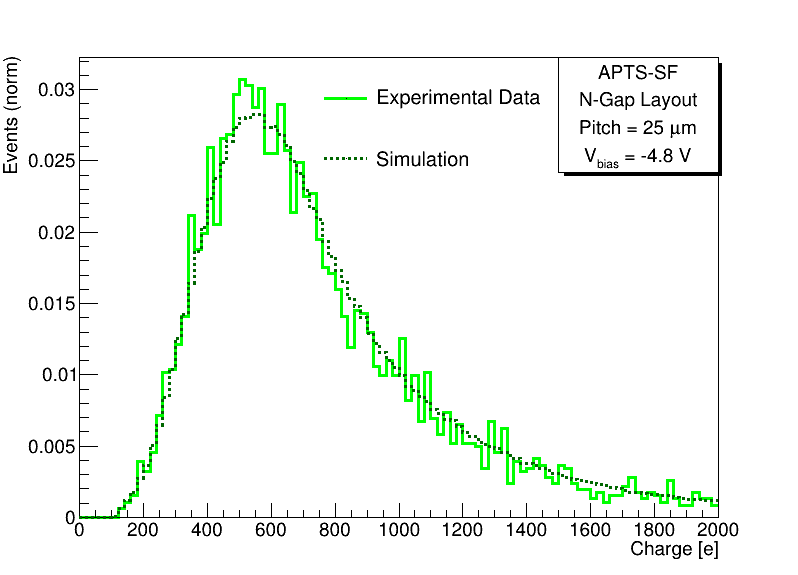}	
	\caption{Comparison of experimental data and simulations of charge distribution of the seed pixel for the n-gap layout at -4.8 V bias voltage.} 
	\label{fig:comp_charge_seed_gap_4v8}%
\end{figure}

\begin{figure}[htb]
	\centering 
	\includegraphics[width=0.5\textwidth]{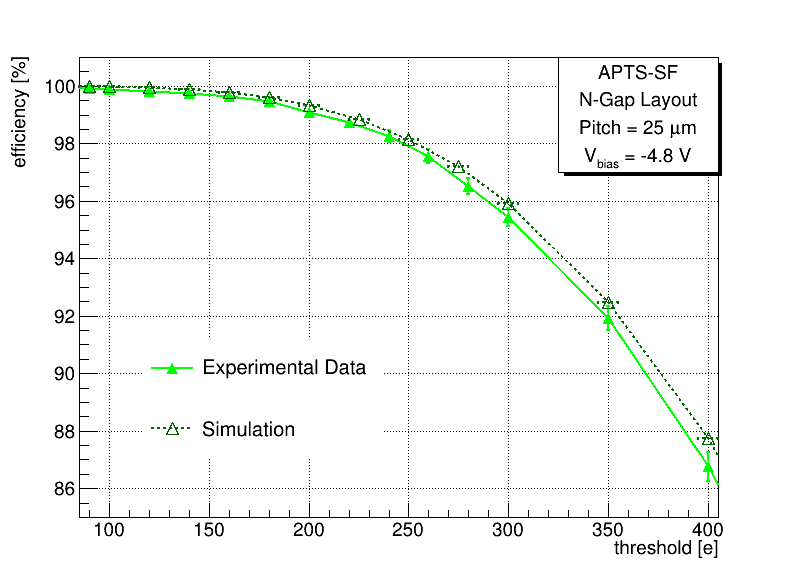}	
	\caption{Comparison of experimental data and simulations of efficiency as a function of the threshold for the n-gap layout at -4.8 V bias voltage. Only statistical uncertainties are included.} 
	\label{fig:comp_thScan_eff_gap}%
\end{figure}

Simulations are compared to experimental data in order to validate the employed simulation approach. Figures \ref{fig:comp_charge_seed_gap_4v8} and \ref{fig:comp_thScan_eff_gap} show, respectively, the comparison of the seed pixel charge distribution and the efficiency as a function of the threshold for a detector in the n-gap design. Both plots show a good agreement between simulation and experimental data.

In figure~ \ref{fig:comp_charge_seed_gap_4v8}, both charge distributions follow the trend of a Landau distribution convolved with a Gaussian, representing effects in the interaction of minimum ionizing particles traversing thin sensors, such as stochastic fluctuations of charge deposition and electronic noise. This convolved function is fitted to extract the \textit{most probable value} (MPV) for the charge collection. The obtained MPVs are around 500\,electrons.

Figure \ref{fig:comp_thScan_eff_gap} shows that the agreement between experimental data and simulations for the detector efficiency is within 1\%, only considering statistical uncertainties.

Ongoing studies of a detector in the standard design show a slight disagreement between simulations and experimental data. The different compatibility depending on the sensor layouts can be explained by the susceptibility of the standard layout to effects due to charge mobility dominated by diffusion. This makes the standard design more sensitive to parameters such as carrier lifetimes. Further studies are planned to understand the origin of these differences.

\section{Summary and Conclusions}

Given their low material budget and improved performance parameters, MAPS produced in the TPSCo 65\,nm CMOS imaging technology are a promising proposal for vertex detectors at future lepton colliders.

Different small collection electrode MAPS technology demonstrators have been studied through test-beam characterization and simulations. The efficiency, mean cluster size, and charge distribution have been investigated for the n-gap design. Results show a consistent behavior with the same design studied in a 180\,nm technology; there are no significant differences in the efficiency and mean cluster size for the investigated bias voltages. The overall mean cluster size is small because of the charge-sharing effects constrained by the enlarged depleted volume and the high lateral electric field at the edges.

Simulations have been performed with a combination of TCAD and Monte Carlo frameworks on a sensor structure using generic doping profiles. The device simulations on TCAD produce the complex electric fields characteristic of small collection electrode MAPS. Monte Carlo simulations produce performance parameters that are directly comparable to experimental data. Simulations for the n-gap design exhibit compatibility with experimental data, opening a promising path toward validating them.

The next prototypes under study have been produced in the second submission to the foundry: DESY Chip V2, which includes full in-pixel functionalities, and H2M (hybrid-to-monolithic), including in-pixel analog and digital processing electronics and a pixel matrix of 64 × 16. These prototypes are being investigated using the same methods presented here.

\section*{Acknowledgements}

The TANGERINE project is funded by the Helmholtz Innovation Pool, 2021 - 2024. This work has been sponsored by the Wolfgang Gentner Programme of the German Federal Ministry of Education and Research (grant no. 13E18CHA) and has received funding from the European Union’s Horizon 2020 Research and Innovation programme (GA no. 101004761).

Measurements presented have been performed at the test-beam Facility at DESY Hamburg (Germany), a member of the Helmholtz Association (HGF).

The authors express their gratitude to the CERN EP R\&D WP 1.2 and especially to the APTS designers and the ALICE ITS3 measurement team for their support.

(c) All figures and pictures by the author(s) under a \href{https://creativecommons.org/licenses/by/4.0/}{CC BY 4.0} license, unless otherwise stated.

\bibliographystyle{elsarticle-num}
\bibliography{hstd23}

\begin{thebibliography}{10}
\expandafter\ifx\csname url\endcsname\relax
  \def\url#1{\texttt{#1}}\fi
\expandafter\ifx\csname urlprefix\endcsname\relax\def\urlprefix{URL }\fi
\expandafter\ifx\csname href\endcsname\relax
  \def\href#1#2{#2} \def\path#1{#1}\fi

\bibitem{european_strategy}
{European Strategy Group}, {2020 Update of the European Strategy for Particle Physics} (2020).
\newblock \href {https://doi.org/http://dx.doi.org/10.17181/ESU2020Deliberation} {\path{doi:http://dx.doi.org/10.17181/ESU2020Deliberation}}.

\bibitem{Wennlof2022}
H.~Wennlöf, et~al., {The Tangerine project: Development of high-resolution 65 nm silicon MAPS}, Nucl. Instrum. Methods Phys. Res. A 1039 (2022) 167025.
\newblock \href {https://doi.org/10.1016/j.nima.2022.167025} {\path{doi:10.1016/j.nima.2022.167025}}.

\bibitem{Feindt2022}
F.~Feindt, et~al., {Towards a new generation of Monolithic Active Pixel Sensors}, Nucl. Instrum. Methods Phys. Res. A 1047 (2023) 167821.
\newblock \href {https://doi.org/https://doi.org/10.1016/j.nima.2022.167821} {\path{doi:https://doi.org/10.1016/j.nima.2022.167821}}.

\bibitem{Simancas2023}
A.~Simancas, et~al., {Developing a Monolithic Silicon Sensor in a 65nm CMOS Imaging Technology for Future Lepton Collider Vertex Detectors}, in: 2022 IEEE (NSS/MIC), 2022, pp. 1--7.
\newblock \href {https://doi.org/10.1109/NSS/MIC44845.2022.10398964} {\path{doi:10.1109/NSS/MIC44845.2022.10398964}}.

\bibitem{EPreport}
{CERN Collaboration}, \href{http://cds.cern.ch/record/2764386?ln=en}{{Strategic R\&D Programme on Technologies for Future Experiments - Annual report 2020}}, CERN Document Server (2020).
\newline\urlprefix\url{http://cds.cern.ch/record/2764386?ln=en}

\bibitem{alice}
{ALICE Collaboration}, {Letter of Intent for an ALICE ITS Upgrade in LS3} (2019).

\bibitem{clic_CDR}
L.~Linssen, CERN, {Physics and detectors at CLIC: CLIC conceptual design report}, CERN.

\bibitem{Deng2023}
W.~Deng, et~al., {Design of an analog monolithic pixel sensor prototype in TPSCo 65 nm CMOS imaging technology}, J. Instrum. 18 (1 2023).
\newblock \href {https://doi.org/10.1088/1748-0221/18/01/C01065} {\path{doi:10.1088/1748-0221/18/01/C01065}}.

\bibitem{alice_apts}
F.~Carnesecchi, et~al., {Characterisation of Analogue Monolithic Active Pixel}, Submitted to Nucl. Instrum. Methods Phys. Res. A (2024).

\bibitem{Senyukov2013}
S.~Senyukov, et~al., {Charged particle detection performances of CMOS pixel sensors produced in a 0.18µm process with a high resistivity epitaxial layer}, Nucl. Instrum. Methods Phys. Res. A 730 (2013) 115--118.
\newblock \href {https://doi.org/10.1016/j.nima.2013.03.017} {\path{doi:10.1016/j.nima.2013.03.017}}.

\bibitem{Snoeys2017}
W.~Snoeys, et~al., {A process modification for CMOS monolithic active pixel sensors for enhanced depletion, timing performance and radiation tolerance}, Nucl. Instrum. Methods Phys. Res. A 871 (2017) 90--96.
\newblock \href {https://doi.org/10.1016/j.nima.2017.07.046} {\path{doi:10.1016/j.nima.2017.07.046}}.

\bibitem{Muenker_thesis}
M.~Munker, \href{http://hss.ulb.uni-bonn.de/diss_online}{{Test beam and simulation studies on High Resistivity CMOS pixel sensors}} (2018).
\newline\urlprefix\url{http://hss.ulb.uni-bonn.de/diss_online}

\bibitem{Munker2019}
M.~Munker, et~al., {Simulations of CMOS pixel sensors with a small collection electrode, improved for a faster charge collection and increased radiation tolerance}, J. Instrum. 14 (2019).
\newblock \href {https://doi.org/10.1088/1748-0221/14/05/C05013} {\path{doi:10.1088/1748-0221/14/05/C05013}}.

\bibitem{Snoeys_pixel22}
W.~Snoeys, et~al., Optimization of a 65 nm cmos imaging process for monolithic cmos sensors for high energy physics, PoS, 2022, p.~83.
\newblock \href {https://doi.org/10.22323/1.420.0083} {\path{doi:10.22323/1.420.0083}}.

\bibitem{caribou1}
H.~Liu, et~al., {Development of a modular test system for the silicon sensor R\&D of the ATLAS Upgrade}, J. Instrum. 12 (2017).
\newblock \href {https://doi.org/10.1088/1748-0221/12/01/P01008} {\path{doi:10.1088/1748-0221/12/01/P01008}}.

\bibitem{caribou2}
T.~Vanat, {Caribou – A versatile data acquisition system}, Vol. 370, Sissa Medialab, 2020, p. 100.
\newblock \href {https://doi.org/10.22323/1.370.0100} {\path{doi:10.22323/1.370.0100}}.

\bibitem{mimosa}
H.~Jansen, et~al., {Performance of the EUDET-type beam telescopes}, Eur. Phys. J. TI 3 (2016) 1--18.
\newblock \href {https://doi.org/10.1140/epjti/s40485-016-0033-2} {\path{doi:10.1140/epjti/s40485-016-0033-2}}.

\bibitem{telepix}
L.~Huth, et~al., {Upgrading the beam telescopes at the DESY II Test Beam Facility}, Nucl. Instrum. Methods Phys. Res. A 1040 (2022) 167183.
\newblock \href {https://doi.org/10.1016/j.nima.2022.167183} {\path{doi:10.1016/j.nima.2022.167183}}.

\bibitem{tlu}
P.~Baesso, et~al., {The AIDA-2020 TLU: A flexible trigger logic unit for test beam facilities}, J. Instrum. 14 (2019).
\newblock \href {https://doi.org/10.1088/1748-0221/14/09/P09019} {\path{doi:10.1088/1748-0221/14/09/P09019}}.

\bibitem{eudaq2}
Y.~Liu, et~al., {EUDAQ2 - A flexible data acquisition software framework for common test beams}, J. Instrum. 14 (10 2019).
\newblock \href {https://doi.org/10.1088/1748-0221/14/10/P10033} {\path{doi:10.1088/1748-0221/14/10/P10033}}.

\bibitem{corry}
D.~Dannheim, et~al., {Corryvreckan: a modular 4D track reconstruction and analysis software for test beam data}, J. Instrum. 16 (2021) P03008.
\newblock \href {https://doi.org/10.1088/1748-0221/16/03/P03008} {\path{doi:10.1088/1748-0221/16/03/P03008}}.

\bibitem{desy-tb}
R.~Diener, et~al., {The DESY II test beam facility}, Nucl. Instrum. Methods Phys. Res. A 922 (2019) 265--286.
\newblock \href {https://doi.org/10.1016/j.nima.2018.11.133} {\path{doi:10.1016/j.nima.2018.11.133}}.

\bibitem{gbl}
C.~Kleinwort, \href{https://gitlab.desy.de/claus.kleinwort/general-broken-lines}{{General Broken Lines - GitLab}}.
\newline\urlprefix\url{https://gitlab.desy.de/claus.kleinwort/general-broken-lines}

\bibitem{tcad+mc}
D.~Dannheim, et~al., {Combining TCAD and Monte Carlo methods to simulate CMOS pixel sensors with a small collection electrode using the Allpix2 framework}, Nucl. Instrum. Methods Phys. Res. A 964 (2020) 163784.
\newblock \href {https://doi.org/10.1016/j.nima.2020.163784} {\path{doi:10.1016/j.nima.2020.163784}}.

\bibitem{Spannagel2018}
S.~Spannagel, et~al., {Allpix2: A modular simulation framework for silicon detectors}, Nucl. Instrum. Methods Phys. Res. A 901 (2018) 164--172.
\newblock \href {https://doi.org/10.1016/j.nima.2018.06.020} {\path{doi:10.1016/j.nima.2018.06.020}}.

\end{thebibliography}






\end{document}